# Enhanced magnetic, electrical, and magnetostrictive properties of La-doped SrCoO$_3$ synthesized by microwave heating

L. A. Longchar, M. Manikandan, and R. Mahendiran*

*Department of Physics, 2 Science Drive 3,*
*National University of Singapore, Republic of Singapore, Singapore-117551*

**Abstract**

We report microwave-assisted synthesis and physical properties of La-doped SrCoO$_{3-\delta}$. The Sr$_{0.8}$La$_{0.2}$CoO$_{3-\delta}$ synthesized via microwave heating (MWH) exhibits superior physical properties compared to a nominally identical composition obtained by conventional heating (CH) in an electrical furnace. The MWH sample exhibits an enhanced ferromagnetic Curie temperature (158 K in CH and 176 K in MWH samples) and higher saturation magnetization but a smaller coercive field and magnetoresistance at 10 K compared to the CH sample. While the dc resistivity increases substantially below 120 K in the CH sample, it is metallic from 350 to 10 K in the MWH sample. Further, the longitudinal magnetostriction of the MWH sample ($\lambda_{par}$ = 247 ppm for $H$ = 50 kOe) at 10 K is also higher than that of the CH sample ($\lambda_{par}$ = 128 ppm). The observed enhanced properties of the MWH sample can not be attributed to grain size and grain boundaries but are likely to arise from lesser oxygen defects and partial oxidation of Co$^{3+}$ into Co$^{4+}$, however, the exact mechanism is an open question.

*Corresponding author: phyrm@nus.edu.sg*

The perovskite cobaltite $La_{1-x}Sr_xCoO_3$ exhibits a rich variety of physical phenomena such as ferromagnetism and metallicity for $0.25 \leq x \leq 0.5$[1,2,3], temperature induced spin-state transition of $Co^{3+}$ ion in pure undoped $LaCoO_3$[4], giant magnetoresistance in cluster-glass insulating compositions ($0.09 \leq x \leq 0.18$)[5,6,7], anomalous magnetovolume expansion above 600 kOe in $x = 0.0$ due to magnetic field induced low-spin ($LS$, spin $S = 0$) to high-spin ($HS$, $S = 2$) state transition of $Co^{3+}$ ion[8,9], but anisotropic magnetostriction in $x = 0.3$ and $0.5$[10,11], magnetic shape memory effect in $x = 0.2$[12], anomalous Nernst effects in single crystals[13] and polycrystals[14], strain-driven metal to insulator transition in thin films[15], etc. The substitution of $Sr^{2+}$ for $La^{3+}$ in the non-magnetic insulator $LaCoO_3$ oxidizes $Co^{3+}(d^6)$ ions into $Co^{4+}(d^5)$ in proportion to the $Sr^{2+}$ content ($x$), which amounts to hole doping. It was assumed that mixed valence of Co ions ($Co^{3+}$ and $Co^{4+}$) was essential for ferromagnetism[2,4] in analogy to the hole-doped manganite $La_{1-x}Sr_xMnO_3$ wherein doped holes mediate ferromagnetic interactions between $Mn^{3+}$ and $Mn^{4+}$ ions by Zener's double exchange interaction. However, the occurrence of ferromagnetism and metallicity in nearly oxygen stoichiometric $SrCoO_3$ single crystal with only $Co^{4+}$ ions excludes the need for mixed valence and double exchange interaction[16]. But, oxygen deficiency in $La_{1-x}Sr_xCoO_3$ series increases with $x > 0.5$[17,18]. Synthesis of $SrCoO_3$ in air and ambient pressure yields brownmillerite $Sr_2Co_2O_5$ phase which is antiferromagnetic and insulating. Preparation of bulk $SrCoO_{3-\delta}$ ($\delta < 0.5$) requires post annealing under high oxygen pressure ($P \sim 10\text{-}150$ MPa) for a week[16,17] or synthesis at high pressure ($\sim 6$ to $8$ GPa) using a diamond anvil cell with pre or post oxygen annealing[19,20]. The maximum Curie temperature of $T_c = 305$ K is realized in a high-pressure synthesized $SrCoO_3$ single crystal[16], whereas $T_c$ varies between 220 K and 280 K in polycrystalline samples/thin films depending on the oxygen partial pressure[17,21,22,23]. Alternatively, electrochemical oxidation increases the $T_c$ in Sr-rich compositions ($x \geq 0.5$) without applying a high-pressure but the oxidation process takes several days ($> 150$ hours)[24,25,26,27,28,29]. The obtained bulk samples are mechanically soft to measure magnetomechanical properties like magnetostriction.

Here, we report that the rapid synthesis ($\sim 30$ min) and sintering of $Sr_{0.8}La_{0.2}CoO_{3-\delta}$ under microwave irradiation in air and ambient pressure not only enhances $T_c$, saturation magnetization, magnetoresistance, and magnetostriction but also promotes metallic behaviour in the electrical resistivity. We have recently reported microwave-assisted synthesis of $La_{0.5}Sr_{0.5}CoO_3$, for which $T_c$ and saturation magnetization values were found to be identical to a sample synthesized by conventional heating in an electrical furnace[14]. Unlike the conventional heating (CH) synthesis via solid state reaction in an electrical furnace, wherein well mixed and ground powders of reactants taken in a crucible is heated by infrared radiation from a current carrying heating element and thermal conduction via walls of the crucible, microwave energy is directly converted into thermal energy within the reactants either by reorientation of localized electrical dipoles or motion of ions and they become out of phase with the electric field of microwave radiation in microwave heating (MWH)[30]. Eddy current and magnetic hysteresis losses at microwave frequencies aid heating. The MWH is rapid (~ 10 to 300 °C/min, depending on the efficiency of reactants to absorb microwave radiation) because it occurs at the molecular level and heat propagates from the interior of grains to the surface (volumetric heating). Our attempt to synthesize a single-phase perovskite $SrCoO_3$ by the CH or MWH method was unsuccessful, and $Sr_{0.9}La_{0.1}CoO_{3-\delta}$ pellets were not hard enough for measuring magnetostriction. Hence, we chose $Sr_{0.8}La_{0.2}CoO_{3-\delta}$ to investigate the influence of the MWH on magnetic, electrical, and magnetostriction properties.

Ten grams of $La_2O_3$, $Co_3O_4$ and $SrCO_3$ taken in a stoichiometric ratio were mixed homogeneously using a pestle and mortar. For the synthesis via solid state reaction by CH method, five grams of mixed powder was heated to 1200 °C in an electrical furnace with a heating rate of 4 °C/min, calcined at 1200 °C for 12 hours and cooled to room temperature with a rate of 4 °C/min. The obtained powder was re-ground and pressed into a 10 mm diameter disc using a hydraulic press, and the above heating process was repeated. For the MW synthesis method, a research-grade microwave (MW) furnace (Milestone PYRO muffle furnace) operating at 2.45 GHz was used. Five grams of the mixed powder taken in an alumina crucible was heated to 1200 °C (heating rate of 40 °C/min) by microwave irradiation power of 1600 Watts and soaked at 1200 °C for 30 minutes before switching off the MW power. Cooling started immediately after the power was cut off and the sample cooled to room temperature in 100 minutes. The temperature of the crucible was measured by a pyro sensor installed in the furnace. Again, the mixture was re-ground homogenously, pelletized using a hydraulic press and the pellet was subjected to microwave heating by repeating the above heating procedure. Overall, the sample was at 1200 °C for 24 hours in CH, whereas only for 60 minutes in the MWH method.

The X-ray diffraction patterns of $Sr_{0.8}La_{0.2}CoO_{3-\delta}$ samples prepared via the MWH and CH synthesis routes are shown in Figures 1(a) and (b), respectively. The room temperature diffraction spectra of both samples were refined with a cubic structure ($Pm\overline{3}m$ space group). The resulting fitting patterns confirm that a single phase is formed. The lattice parameter for the MWH sample is $a = 3.8396(1)$ Å with a unit cell volume of 56.6 Å³, while $a = 3.8393(9)$

Å and volume is 56.5 Å³ for the CH sample. The Co-O bond lengths are nearly identical, 1.9198(3) Å for MWH and 1.9197(8) Å for CH, and the Co-O-Co angle is found to be 180° in both samples. Hence, differences in structural parameters are very small between these two samples. Insets in figure 1(a) and (b) show SEM images of the MWH and CH samples, respectively. The grains of the CH sample have a melt-densified nature and fused into larger grains, whereas the MWH sample displays distinct grain boundaries with a mixture of solid grains and tiny fragmentations. The average grain size of the MWH sample is smaller (=3.5 μm) than the CH sample (=15.5 μm).

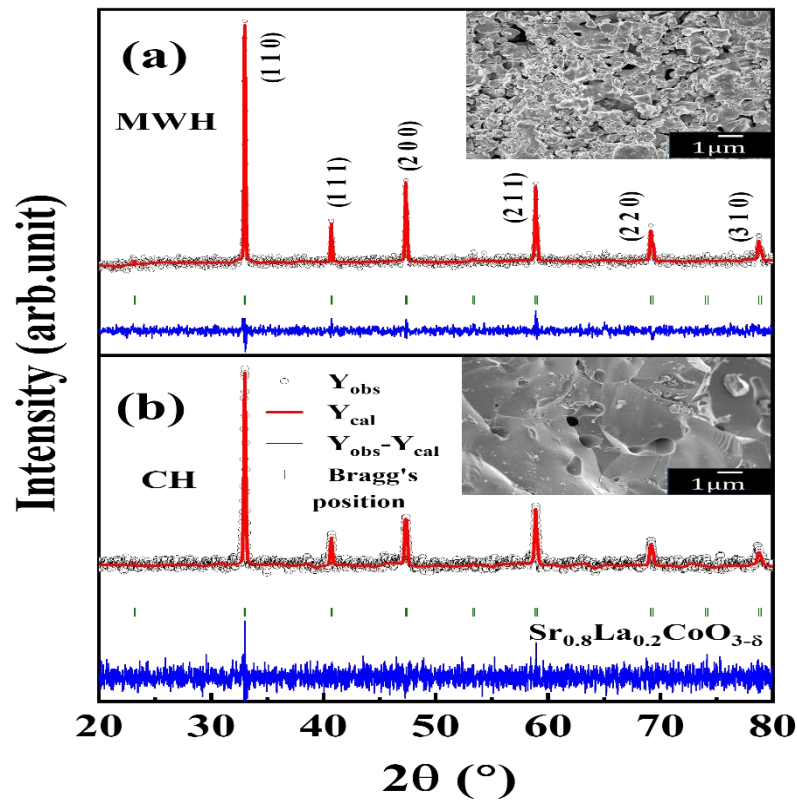

**FIG. 1.** Comparison of X-ray diffraction patterns for $Sr_{0.8}La_{0.2}CoO_{3-\delta}$ samples synthesized by **(a)** microwave heating (MWH) and **(b)** Conventional Heating (CH) methods. The black open circles represent the raw data and the red continuous lines are fits to the data. Blue lines are the difference between the data and fit. Insets: Scanning Electron Microscope images at 1 μm scale.

The main panel of Figure 2(a) illustrates the temperature dependence of magnetization, $M(T)$, of the MWH and CH samples upon cooling from 350 K under a magnetic field of $H = 1$ kOe. The *M(H)* at 10 K for both samples is compared in the inset. The rapid rise of *M(T)* signals a transition from paramagnetic to ferromagnetic state. The Curie temperature ($T_c$) is identified from the inflection point of *dM/dT* in each curve. The MWH sample ($T_c$ = 176 K) is 18 K higher than that of the CH sample ($T_c$ = 158 K). The saturation magnetization of the MWH sample ($M_s$ = 1.32 $\mu_B$/f.u.) is also higher than the CH sample ($M_s$ = 1.22 $\mu_B$/f.u.), but the coercive field ($H_c$ = 1200 Oe) is smaller than the CH sample ($H_c$ = 5280 Oe). The larger $H_c$ suggests a stronger pinning of domain walls by oxygen defects in CH sample.

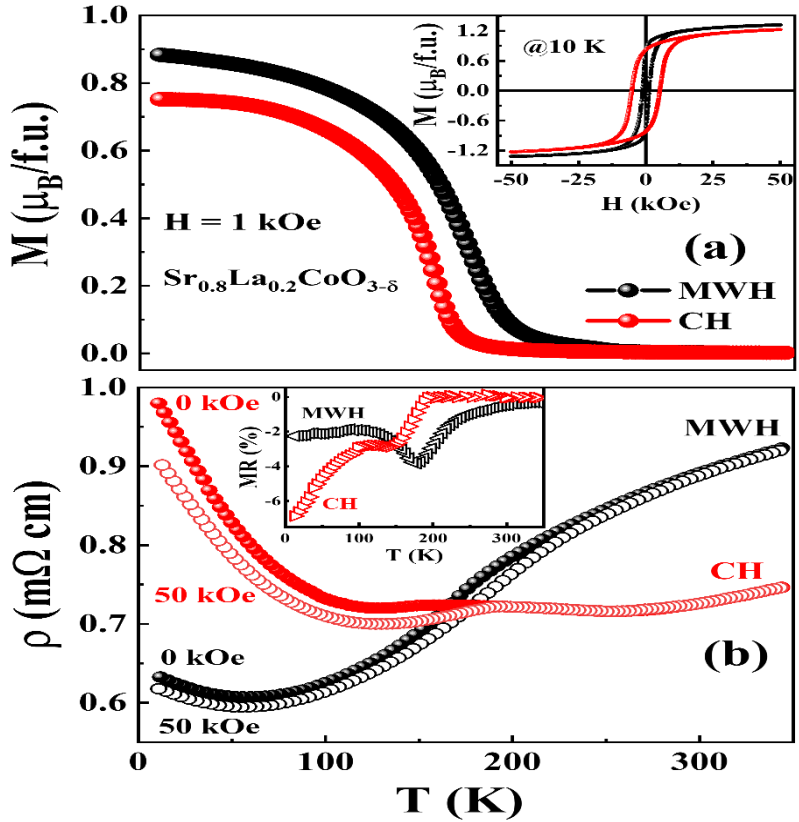

**FIG. 2. (a)** Comparison of *M(T)* in a magnetic field of *H* = 1 kOe for $Sr_{0.8}La_{0.2}CoO_{3-\delta}$ samples synthesized by CH and MWH. The inset shows the magnetic hysteresis loops recorded at 10 K in both samples. **(b)** *ρ(T)* in the MWH and CH samples under *H* = 0 and 50 kOe. The inset compares the temperature dependence of magnetoresistance (*MR*) in both samples derived from the temperature sweeps.

Figure 2(b) compares the temperature dependence of resistivity *ρ(T)* of MWH and CH samples in a magnetic field *H* = 0 and 50 kOe. The inset compares the temperature dependence of magnetoresistance (*MR*) in both samples for *H* = 50 kOe. *ρ(T)* of MWH sample exhibits metallic behaviour (*dρ/dT* > 0) in both paramagnetic and ferromagnetic regions with a slope change around $T_c$ due to a decrease in spin-disorder scattering of conduction electrons. There is a slight upturn below 50 K that was seen earlier in La-rich polycrystalline cobaltites/thin films and attributed to weak localization or electron-electron interaction[4,5,6]. The applied magnetic field smears the slope change around $T_c$ and reduces the resistivity, resulting in negative magnetoresistance. On the other hand, $\rho(T)$ of the CH sample decreases with temperature from $T$ = 350 K to 265 K, followed by a weak temperature dependence between $T$ = 265 K - 130 K and a rapid increase below 120 K. This rapid increase below 120 K is likely due to trapping of charge carriers at oxygen defects. At 10 K, $\rho$ = 0.63 m$\Omega$ cm in the MWH sample which is comparable to the CH sample for which $\rho$= 0.98 m$\Omega$ cm. As the temperature is increased from 10 K, the magnitude of magnetoresistance (*MR*) in the MWH sample increases and reaches a maximum (= 4 %) at $T_c$, then drops to a negligibly small value above $T_c$. However, the *MR* in CH sample is largest (= 6.8 %) at 10 K and does not show a peak around $T_c$, unlike in the MWH sample.

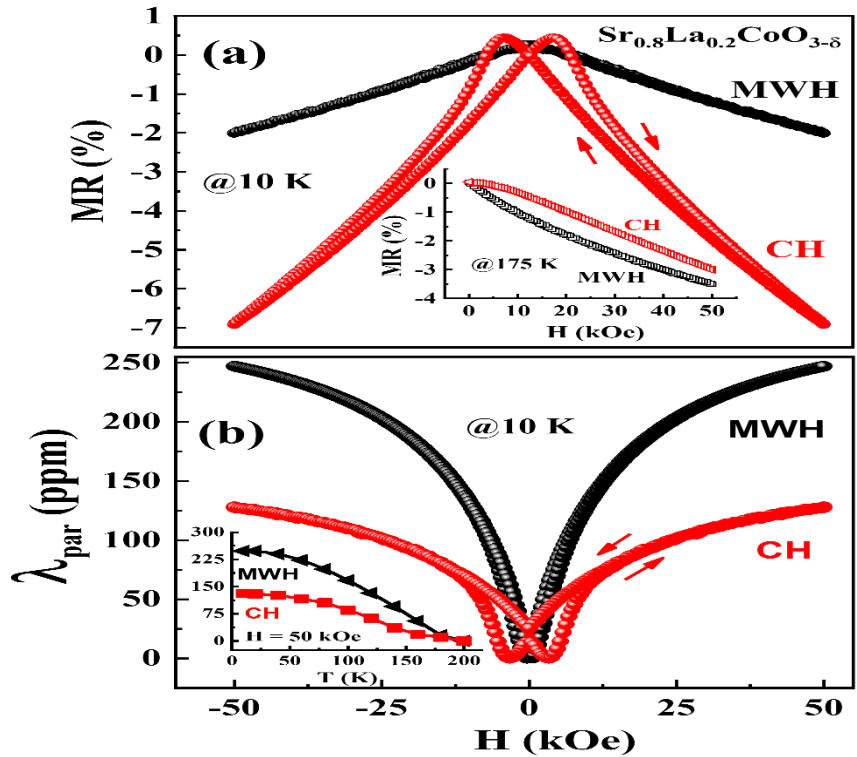

**FIG. 3.** Magnetic field dependence of **(a)** Magnetoresistance (*MR*) and **(b)** Longitudinal magnetostriction ($\lambda_{par}$) at 10 K in MWH and CH samples of $Sr_{0.8}La_{0.2}CoO_{3-\delta}$. Insets: comparison

of **(a)** MR as a function of $H$ at 175 K and **(b)** temperature dependence of $\lambda_{par}$ obtained from the field sweep data at different temperatures, in MWH and CH samples.

The main panel of Figure 3(a) contrasts the field dependence of $MR$ at 10 K in both samples and the inset compares it at 175 K. The $MR$ in CH sample exhibits positive peaks around $\pm H_c$ and hysteresis over a wider field range compared to MWH sample. The $MR$ is negative above $H_c$ and its magnitude increases monotonically with the field up to $H = 50$ kOe without showing a saturation. The maximum $|MR| \approx 7$ % (2 %) in the CH (MW) sample. The field dependence of $MR$ is distinct in both samples at 175 K: It is concave at low fields in MWH sample but convex in CH sample, which reflects that the MWH sample is in the FM state whereas the CH sample is in the paramagnetic state at 175 K.

The longitudinal magnetostriction, $\lambda_{par} = [L(H)-L(0)]/L(0)$, where $L(H)$ is the sample length measured along the applied field (H) direction and $L(0)$ is the initial length without the field at 10 K also shows remarkable differences between these two samples (Figure. 3(b)). $\lambda_{par}$ exhibits pronounced hysteresis in the field range -20 kOe $\leq H \leq$ +20 kOe and minima around $\pm H_c$ in the CH sample and increases smoothly up to the maximum field. At $H = 50$ kOe, $\lambda_{par} = 247$ ppm in the MWH sample whereas it is nearly half ($\lambda_{par} = 128$ ppm) in the CH sample. As the temperature increases from 10 K, $\lambda_{par}$ decreases and becomes negligible above $T_c$ in both samples as shown in the inset of figure 3(b).

The results presented above indicate that the microwave heating has facilitated an increase of $T_c$, $M_s$, $\lambda_{par}$, and reduced the $H_c$ and the magnitude of $MR$ at 10 K. The MWH also reduced the value of resistivity and rendered it metallic from 350 K to 10 K barring a small upturn around 50 K. Since the average grain size of the MWH sample is much smaller than that of the CH sample, it is expected that the MWH sample should show higher resistivity due to scattering of charges at grain boundaries rather than a decrease. These results suggest that factors other than microstructure play a role. We suspect that the oxygen deficiency is less in the MWH sample compared to the CH sample. According to Sunstrom et al.[25] $T_c$ increased from 227 K for $Sr_{0.8}La_{0.2}CoO_{3-\delta}$ sample prepared in air via a soft-chemical route (Pechini gel method) to 272 K for the electrochemically oxidized sample V. Pralong et al.[26] found $T_c$ to increase from 170 K for $Sr_{0.8}La_{0.2}CoO_{2.75}$ (synthesized by the sol-gel technique and cooled in air at a rate of 1°C/min) to 290 K for $Sr_{0.8}La_{0.2}CoO_3$ (fully oxidized by the electrochemical method) but reduced to $T_c \sim 70$ K in $Sr_{0.8}La_{0.2}CoO_{2.65}$. The fully oxidized $Sr_{0.8}La_{0.2}CoO_3$ also showed higher $M_s$ (= 1.9 $\mu_B$/f.u at 50 kOe.) compared to $M_s = 0.45$ $\mu_B$/f.u. in $Sr_{0.8}La_{0.2}CoO_{2.75}$ ($M_s$ = 0.45 $\mu_B$/f.u.). The observed value of $M_{50kOe}$ = 1.32 $\mu_B$/f.u. in our MWH sample is higher than $Sr_{0.8}La_{0.2}CoO_{2.75}$ but closer to ~1.5 $\mu_B$/f.u. found in $Sr_{0.8}La_{0.2}CoO_{2.85}$. So, our MWH sample synthesized in air and ambient pressure is likely to have lesser oxygen defects compared to the CH sample and the oxygen content is likely to be between 2.80 and 2.85 in the MWH sample and below 2.8 in the CH sample. Although the observed increase of $T_c$ via the microwave-assisted synthesis is modest compared to electrochemical oxidation, the sample was synthesized in a fraction of the time in air without using an aqueous electrolyte and it was mechanically hard. The increased $T_c$ in the MWH sample can be attributed to an increase in exchange integral ($J_{ex}$) due to enhanced overlap of O-$2p$ and Co-$3d$ orbitals following the increase in oxygen content. An increase in oxygen content also means an increase in $Co^{4+}$ ($t_{2g}^4 e_g^1$) content[25]. Unfortunately, the increase in oxygen content is small in our sample to be detected by changes in Co-O bond length or Co-O-Co bond angle using the laboratory X-ray diffractometer. Investigation using synchrotron X-ray diffraction will give a better insight.

The mechanism by which MWH increases oxygen content is not fully understood at present. Microwave irradiation seems to oxidize a fraction of LS-$Co^{3+}$ into IS-$Co^{4+}$ ions by absorbing oxygen from air. We would like to draw reader's attention to the fact that an increase in $T_c$ and a decrease in saturation magnetization was observed in spinel $NiMn_2O_4$ under microwave magnetic field heating in a single mode microwave cavity[31]. Although no structural differences were detected between the samples synthesized by conventional heating and microwave magnetic field irradiation, distinguishable features were seen in the electron spin resonance spectra. It was suggested that the microwave magnetic field promoted partial reduction of $Mn^{3+}$ into $Mn^{2+}$ in the tetrahedral sublattice. Microwave heating is rapid, taking 30 minutes to reach 1200 °C, in contrast to the 12 hours required by the CH method. The cooling is also swift, as the sample is nearly quenched from 1200 °C to 27 °C in 100 minutes, beginning immediately after the MW power is switched off. The cooling rate (~12 °C/min) is much higher than 1-5 °C/min typically used in an electrical furnace. Since the cooling rate in the MW furnace is not controllable, we varied the ramping time, $RT$ = 20, 30, 40 and 50 min to reach 1200 °C but fixed the dwelling time to 30 min at 1200 °C. The top panel of Figure 4(a) shows $M(T)$ and its inset shows $M(H)$ at 10 K for all three $RT$. The middle graph depicts $dM/dT$. We notice that $T_c$ increases from 170 K for $RT$ = 20 min, to 176 K for $RT$ = 30 min, to 179 K for $RT$ = 40 min. The $T_c$ for $RT$ = 50 min sample was the same as that of 40 min sample and hence not shown here. The $M_s$ at 50 kOe also increases from 1.22 to 1.32 to 1.50 $\mu_B$/f.u. for $RT$ = 20, 30, and 40 min, respectively. Thus, the ramping time of 40 minutes seems to be optimum to get the maximum $T_c$ which implies that oxygen intake from the air reaches saturation. The electronic state of Co ions in $Sr_{1-x}La_xCoO_3$ can be described as $t_{2g}^{5-n}e_g^n$ where $n$ is the number of electrons in the $e_g$ band[24]. At the Fermi energy, spin up $\sigma^*$- $e_g$ band overlaps with spin down $\pi^*$- $t_{2g}$ band. Conduction electrons in $\sigma^*$ band magnetically couple with the quasi-localized electrons in the $\pi^*$ band. Increasing oxygen content populates the spin-up $\sigma^*$ band electrons, and elevates the magnetization value. It is to be noted that maximum $T_c$ (= 305 K) and $M_s$ (= 2.5 $\mu_B$/Co) were obtained for single crystalline $SrCoO_3$ where $n$ = 1, i.e., $Co^{4+}$ ion is in the intermediate spin state ($S$ = 3/2).

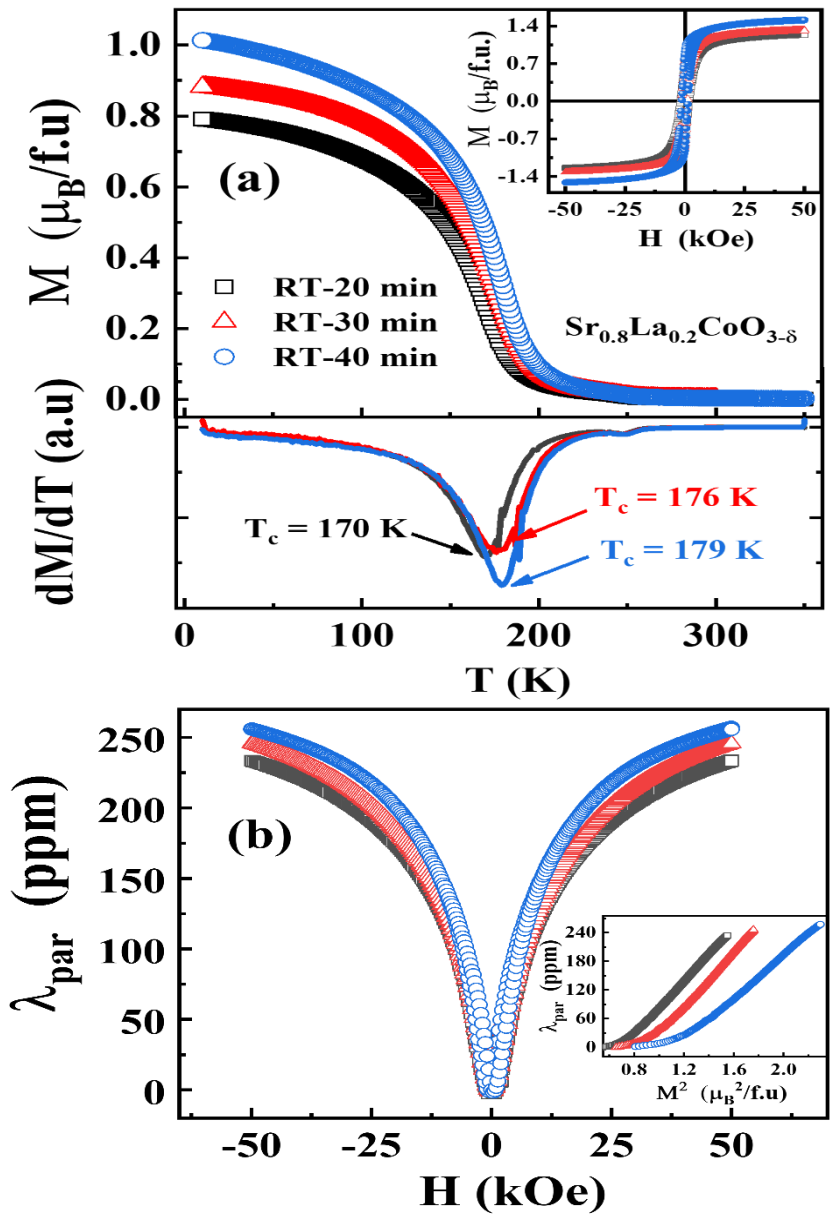

**FIG. 4. (a)** The $M(T)$ of $La_{0.2}Sr_{0.8}CoO_{3-\delta}$ samples synthesized with ramping time of $RT$ = 20, 30, and 40 min. Insets: Comparison of $M(H)$ at 10 K for three ramping time. The bottom figure shows $dM/dT$ vs $T$ with $T_C$ marked by arrows. **(b)** Field dependence of longitudinal magnetostriction ($\lambda_{par}$) at 10 K for the three samples with different ramping times. The inset shows $\lambda_{par}$ versus $M^2$.

The magnetostriction value also increases with $RT$ (see Figure. 4(b)). The $RT$ = 40 min sample shows the maximum $\lambda_{par}$ of 257 ppm, followed by 247 ppm for $RT$ = 30 min and 233 ppm for $RT$ = 20 min. Magnetostriction varies quadratically with magnetization at high fields, as shown in the inset. Earlier magnetostriction studies in the $La_{1-x}Sr_xCoO_{3-\delta}$ series were confined to the La-rich compounds ($x$ = 0.2 - 0.5)[9,10,11]. Although magnetostriction in our MWH sample ($\lambda_{par}$ = 257 ppm for $H$ = 50 kOe at 10 K) is smaller than $\lambda_{par}$ = 400 ppm found in $x$ = 0.5 for the same field[9], it is about four times larger ($\lambda_{par}$ ~ 40 - 60 ppm at 10 K) than in ferromagnetic manganites and it does not saturate unlike in manganites[32]. The large $\lambda_{par}$ values observed in the La-rich compositions could not be understood by conventional spin-orbit interaction because the orbital moment of $Co^{3+}$ ion is usually quenched by the crystalline electrical field of ligands in the octahedral surrounding. It was assumed that ferromagnetic clusters having IS-$Co^{3+}$ ($t_{2g}^5 e_g^1$, spin $S$ = 3/2) and LS-$Co^{4+}$ ($t_{2g}^5$, $S$ = 1/2) coexist with a non-magnetic phase with LS-$Co^{3+}$ ($t_{2g}^6$, $S$ = 0). It was proposed that the applied magnetic field induced the LS to IS state transition of $Co^{3+}$ ions in the non-magnetic region[9]. The double degenerate $t_{2g}$ orbitals of IS $Co^{3+}$ with non-zero orbital angular momentum couple to spin, creating intra-atomic spin-orbit interaction, which deforms the lattice (stretches the length)[9]. A similar mechanism possibly exists in the Sr-rich sample, but the magnitude is small because the LS-$Co^{3+}$ concentration is lower than in La-rich compositions.

In summary, the Sr-rich cubic $Sr_{0.8}La_{0.2}CoO_{3-\delta}$ sample synthesized by microwave heating shows higher Curie temperature, magnetization, magnetostriction, and magnetoresistance at 10 K compared to the sample synthesized in an electrical furnace. It is suggested that microwave-irradiation enabled partial oxidation of $Co^{3+}$ into $Co^{4+}$ by oxygen absorption from air. The exact mechanism behind this process remains an open question. In situ spectroscopic characterizations during microwave heating will be helpful to understand the mechanisms involved. It will be interesting to investigate how much Curie temperature and magnetization can be enhanced if the microwave synthesis is carried out under different oxygen partial pressures. Additionally, exploring magnetostriction in near-stoichiometric $SrCoO_3$ is essential to understand the role of the intermediate spin $Co^{4+}$ in magnetostriction

**Acknowledgements**: R. M acknowledges the Ministry of Education, Singapore for supporting this work (Grant Numbers: WBS: A-8000924-00-00 and A-8000462-00-00).

**AUTHOR DECLARATIONS**

**Credit :L.A. Longchar:** Sample preparation and characterizations, **M. M**: Sample preparation and characterizations. **R. M**: conceptualization, resources, supervision. All three authors discussed and collectively wrote the manuscript.

Data Availability statement: Data collected with the figures used in the manuscript is available from the authors upon reasonable request.

**Conflict of Interest:** The authors have no conflicts to disclose.

References

[1] G. H. Jonker and J. H. Van Santen, Magnetic compounds with perovskite structure III. ferromagnetic compounds of cobalt, Physica **19**, 120 (1953).

[2] P. M. Raccah and J. B. Goodenough, A Localized-Electron to Collective-Electron Transition in the System (La, Sr)CoO$_3$, J. Appl. Phys. **39**, 1209 (1968).

[3] V. G. Bhide, D. S. Rajoria, C. N. R. Rao, G. R. Rao, and V. G. Jadhao, Itinerant-electron ferromagnetism in La$_{1-x}$Sr$_x$CoO$_3$: A Mossbauer study, Phys. Rev. B **12**, 2832 (1975).

[4] M. A. Senaris-Rodriguez and J. B. Goodenough, Magnetic and transport properties of the system La$_{1-x}$Sr$_x$CoO$_3$ (0<$x$≤0.5), J. Solid State Chem. **118**, 323 (1995).

[5] R. Mahendiran and A. K. Raychaudhuri, Magnetoresistance of the spin-state transition compounds La$_{1-x}$Sr$_x$CoO$_3$, Phys. Rev. B **54**, 16044 (1996).

[6] J. Wu and C. Leighton, Glassy ferromagnetism and magnetic phase separation in La$_{1-x}$Sr$_x$CoO$_3$, Phys. Rev. B **76**, 174408 (2003).

[7] D. Samal and P. S. Anil Kumar, Critical re-examination and phase diagram of La$_{1-x}$Sr$_x$CoO$_3$, J. Phys. Cond. Mater., **23**, 106001 (2011).

[8] M. Rotter, Z.-S. Wang, A. T. Boothroyd, D. Prabhakaran, A. Tanaka and M. Doer, Mechanism of spin crossover in LaCoO$_3$ resolved by shape and magnetostriction in pulsed magnetic fields. Sci. Rep. **4**, 7003 (2014).

[9] M. M. Altarawneh, G.-W. Chen, N. Harrison, C. D. Batista, A. Uchida, M. Jaime, D. G. Rickel, S. A. Crooker, C. H. Mielke, J. B. Betts, J. F. Mitchell, and M. J. R. Hoch, Cascade of magnetic field induced spin state transition in LaCoO$_3$, Phys. Rev. Lett. **109**, 037201 (2012).

[10] M. R. Ibarra, R. Mahendiran, C. Marquina, B. García-Landa, and J. Blasco, Huge anisotropic magnetostriction in La$_{1-x}$Sr$_x$CoO$_{3-\delta}$ ($x$ >~ 0.3): field induced orbital instability, Phys. Rev. B **57**, R3217 (1998).

[11] B. Kundys and H. Szymczak, Magnetostriction thin films of cobaltites and manganites, Phys. Status Solidi (a), **201**, 3247 (2004).

[12] A. Yokosuka, H. Kumagai, M. Fukuda, K. Ando, Y. Hara, and K. Sato, Room temperature magnetic shape memory effect in strontium-doped lanthanum cobaltite single crystal, AIP Adv. **10**, 095217 (2020).

[13] T. Miyasato, N. Abe, T. Fujii, A. Asamitsu, S. Onoda, Y. Onose, N. Nagaosa, and Y. Tokura, Crossover behavior of the anomalous Hall effect and anomalous Nernst effect in itinerant ferromagnets, Phys. Rev. Lett. **99**, 086602 (2007).

[14] M. Manikandan, A. Ghosh, and R. Mahendiran, Anomalous Nernst Effect and giant magnetostriction in microwave synthesized La$_{0.5}$Sr$_{0.5}$CoO$_3$, J. Phys. Chem. C **126**, 1152 (2022).

[15] A. D. Rata, A. Herklotz, K. Nenkov, L. Schultz, and K. Dörr, Strain-induced insulator state in La$_{0.7}$Sr$_{0.3}$CoO$_3$, Phys. Rev. B **76**, 012403 (2007).

[16] Y. Long, Y. Kaneko, S. Ishiwata, Y. Taguchi, and Y. Tokura, Synthesis of cubic SrCoO$_3$ single crystal and its anisotropic magnetic and transport properties, J. Phys.: Condens. Matter **23**, 245601 (2011).

[17] H. Taguchi, M. Shimada, and M. Koizumi, Effect of oxygen vacancy on the magnetic properties in the system SrCoO$_{3-\delta}$ ($0 < \delta < 0.5$), J. Solid State Chem. **29** (1979); Electric properties of ferromagnetic La$_{1-x}$Sr$_x$CoO$_3$ ($0.5 \leq x \leq 0.9$), J. Solid State Chem. **33**, 169 (1980).

[18] S. Kolesnik, B. Dabrowski, J. Mais, M. Majjiga, O. Chmaissem, A. Baszczuk, and J. D. Jorgensen, Tuning magnetic and electronic states by control of oxygen content in lanthanum strontium cobaltites, Phys. Rev. B **73**, 214440 (2006).

[19] S. Balamurugan, M. Xu, and E. Takayama-Muromachi, Magnetic and transport properties of high pressure synthesized perovskite cobalt oxide (Sr$_{1-x}$Ca$_x$)CoO$_3$ ($0 \leq x \leq 0.8$), J. Solid State Chem. **178**, 3431 (2005).

[20] H. Takahashi, M. Onose, Y. Kobayashi, T. Osaka, S. Maeda, A. Muyake, M. Tokunaga, H. Sagamaya, Y. Yamaskai, and S. Ishiwata, Possible helimagnetic order in Co$^{4+}$- containing perovskites Sr$_{1-x}$Ca$_x$CoO$_3$, APL Mater. **10**, 111116 (2022)

[21] H. Watanabe, Y. Yamaguchi, H. Oda, and H. Takei, Magnetic and neutron diffraction study of SrCoO$_{3-x}$, J. Magn. Magn. Mater. **15–18**, 521 (1980).

[22] C. K. Xie, Y.F. Nie, B.O. Wells, J. I. Budnick, W. A. Hines, and B. Dabrowski, Magnetic phase separation in SrCoO$_{3-x}$ ($2.5 \leq x \leq 3.0$), Appl. Phys. Lett. **99**, 052503 (2011).

[23] S. Chowdhury, R. J. Choudhary, and D. M. Phase, Spectroscopic aspects of the magnetic interactions in SrCoO$_{2.75}$ and SrCoO$_3$ thin films, J. Alloys and Compounds **869**, 159296 (2021).

[24] P. Bezdicka, A. Wattiaux, J. C. Grenier, M. P. Pouchard, and P. Hagenmuller, Preparation and characterization of fully stoichiometric SrCoO$_3$ by electrochemical oxidation, Z. Anorg. Allg. Chem. **619**, 7 (1993).

[25] J. E. Sunstrom, K. V. Ramanujachary, M. Greenblatt, and M. Croft, The synthesis and properties of chemically oxidized perovskite La$_{1-x}$Sr$_x$CoO$_{3-d}$ ($0.5 \leq x \leq 0.9$), J. Solid State Chem. **139**, 388 (1998).

[26] V. Pralong, V. Caignaert, S. Hébert, C. Marinescu, B. Raveau, and A. Maignan, Electrochemical oxidation and reduction of the La$_{0.2}$Sr$_{0.8}$CoO$_{3-\delta}$ phases: Control of itinerant ferromagnetism and magnetoresistance, Solid State Ionics **177**, 815 (2006).

[27] F. J. Rueckert, Y. F. Nie, C. Abughayada, S. A. Sabok-Sayr, H. E. Mohottalla, J. I. Budnick, W. A. Hines, B. Dabrowski, and B.O. Wells, Suppression of magnetic phase separation in epitaxial SrCoO$_x$ films, Appl. Phys. Lett **102**, 152402 (2013).

[28] M. Chennabasappa, E. Petit, and O. Toulemonde, Toward oxygen fully stoichiometric La$_{1-x}$Sr$_x$CoO$_3$ ($0.5 \leq x \leq 0.9$) perovskites: Itinerant magnetic mechanism more than double exchange one's, Ceramic Int. **46**, 6067 (2020).

[29] X. Li, H. Wang, Z. Cui, Y. Li, S. Xin, J. Zhou, Y. Long, C. Q. Jin, and J. B. Goodenough, Exceptional oxygen evolution reactivities on $CaCoO_3$ and $SrCoO_3$, Sci. Adv. **5**, eaav6262 (2019).

[30] K. J. Rao, B. Vaidhyanathan, M. Ganguli, and P. A. Ramakrishnan, Synthesis of inorganic solids using microwaves, Chem. Mater. **11**, 882 (1999); J. Prado-Gonjal, R. Schmidt, and E. Moran, Microwave-assisted routes for synthesis of complex functional oxides, Inorganics, **3**, 101 (2015).

[31] H. Goto, J. Fukushima, and H. Takizawa, Control of magnetic properties of $NiMn_2O_4$ by a microwave magnetic field under air, Materials **9**, 169 (2016).

[32] M. R. Ibarra, P. Algarabel, C. Marquina, J. Blasco, and J. Garcia, Large magnetovolume effect in Yttrium doped La-Ca-$MnO_3$ perovskite, Phys. Rev. Lett. **75,** 3541(1995);Yu. Bukhantsev, Ya. M. Mukovskii, and H. Szmczak, Magnetic properties of $(La_{0.8}Ba_{0.2})_{0.93}MnO_3$ single crystal, J. Magn. Magn. Mater. **272-276**, 2053 (2004).